\documentclass{iau} 
\usepackage{graphicx}
\usepackage{natbib}
\usepackage{float}
 \renewcommand{\vec}[1]{\mbox{\boldmath$#1$}}

\title[] %% give here short title %%
{Cycle times of early M dwarf stars: mean field models versus observations}

\author[Manfred K{\"u}ker, G{\"u}nther R{\"u}diger, Katalin Olah, Klaus Strassmeier]   %% give here short author list %%
{Manfred K{\"u}ker$^1$, G{\"u}nther R{\"u}diger$^1$, Katalin Ol\'ah$^2$
 \and Klaus Strassmeier$^1$}

\affiliation{$^1$Leibniz-Institut f\"ur Astrophysik Potsdam, \\ An der Sternwarte 16,
14482 Potsdam, Germany \\ email: {\tt mkueker@aip.de}, {\tt gruediger@aip.de}, {\tt kstrassmeier@aip.de} \\[\affilskip]
$^2$Konkoly Observatory, \\ Budapest
Hungary \\email: {\tt olahkatalin5@gmail.com}}

\pubyear{2019}
\volume{354}  %% insert here IAU Symposium No.
\jname{Solar and Stellar Magnetic Fields: Origins and Manifestations}
\editors{A. Kosovichev, K. Strassmeier \& M. Jardine, eds.}
\begin{document}

\maketitle

\begin{abstract}  
Observations of early-type M stars suggest that there are two characteristic cycle times, one of order one year for fast rotators ($P_{\rm rot} < 1$ day) and another of order four years for slower rotators. For a sample of fast-rotating stars, the equator-to-pole differences of the rotation rates up to 0.03 rad d$^{-1}$ are also known from Kepler data. These findings are well-reproduced by mean field models. These models predict amplitudes of the meridional flow, from which the travel time from pole to equator at the base of the convection zone of early-type M stars can be calculated. 
As these travel times always exceed the observed cycle times, our findings do not support the flux transport dynamo. 
\end{abstract}

\firstsection % if your document starts with a section,
\section{Introduction}
The solar cycle is widely believed to be the result of a flux transport dynamo in which the toroidal magnetic field is advected towards the equator at the bottom of the convection zone by the large-scale meridional flow. The cycle time is then determined by the flow speed \citep{CS95, DC99B, KR01, BE02B}.
In a traditional $\alpha \Omega$ shell dynamo, on the other hand, the cycle time in the linear regime is 
\begin{equation}
\tau_{\rm cyc} \simeq c_{\rm cyc} \frac{R_*D} {\eta_T},
\label{alfom}
\end{equation}
where $R_*$ is the stellar radius, $D$ the thickness of the convective layer, and $\eta_T$ the 
turbulent magnetic diffusivity, and $c_{\rm cyc}$ a scaling factor \citep{R72}. 
For the Sun, $\eta_T \simeq 10^{12} $ cm$^2$ s$^{-1}$ is required to reproduce the cycle time of eleven years. This value agrees with what is found from the decay of large active regions and from cross helicity $\langle \vec{u} \cdot \vec{b}\rangle. $
 Interestingly, Equation  \ref{alfom} does not contain the rotation rate.
 
\begin{figure}
% \vspace*{-2.0 cm}
\begin{center}
\includegraphics[width=6.7cm]{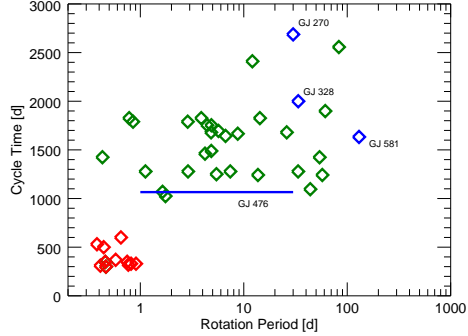} 
% \vspace*{-1.0 cm}
 \caption{Cycle times vs.~rotation period (both in days) for early M stars (diamonds). 
 The red diamonds indicate stars observed by the {\sc Kepler} spacecraft, the blue ones observations with the {\sc Stella} telescope. The blue bar indicates GJ 476, for which the rotation period has not been determined.
 Data from \cite{VK13, VO14, DL16, SR16, WS17, KR19}
 }
   \label{result}
\end{center}
\end{figure}

 As the radii of M dwarfs are smaller than the solar radis,  we expect shorter cycles unless the magnetic diffusivity coefficient is substantially smaller than $10^{12}$cm$^2$s$^{-1}.$  Indeed, cycle times derived from light curves lie in the range from {300\,d} to {2700\,d}, cf.~Fig.~\ref{result} . However, there seems to be a dependence on the rotation period, as very rapidly rotating stars show shorter cycle times.
 
\section{Differential rotation and meridional flow}
\begin{figure}
\begin{center}
\includegraphics[width=11cm]{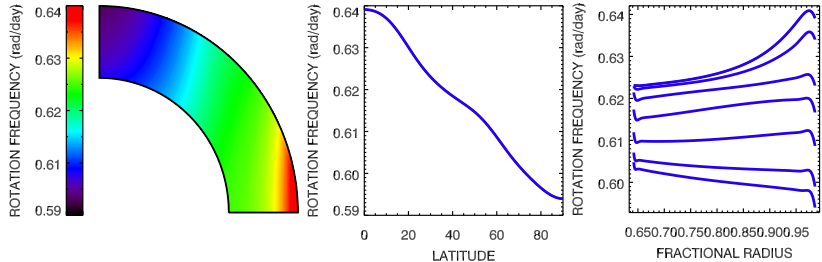}
\caption{Differential rotation of an M star with $M=0.6 M_\odot$ rotating with a period of 10 d. Left: Color contour plot of the angular velocity. Center: Surface rotation as a function of latitude. Right: Angular velocity vs.~fractional stellar radius for different latitudes. From top to bottom: $0^\circ$, $15^\circ$, $30^\circ$, $45^\circ$, $60^\circ$, $75^\circ$, and $90^\circ$.
\label{m06_dr}
}
\end{center}
\end{figure}

Meridional flows are hard to observe even on the Sun. For stars we have to rely on theoretical models.
We have therefore applied the mean field model of \cite{KRK11}, which reproduces the solar differential rotation very well, to a sample of three stellar models. The masses  were chosen to be 0.4, 0.6, and 0.66 solar masses in order  to cover the range of spectral types in the sample of M dwarfs with observed cycle times. The stellar models were computed with the Mesa stellar evolution code of \cite{PB11} assuming solar metallicity. Figure \ref{m06_dr} shows the resulting differential rotation patterns for 0.6 solar masses and a rotation period of ten days. The rotation is solar-type, i.e.~more rapid rotation at the equator than at the poles. However, the isocontours are more cylinder-shaped than what both our model and helioseismology find for the Sun. The latitudinal shear is larger at the surface than at the bottom of the convection zone. The radial shear is positive at the equator and negative at the poles.

\begin{figure}
\begin{center}
\includegraphics[width=11cm]{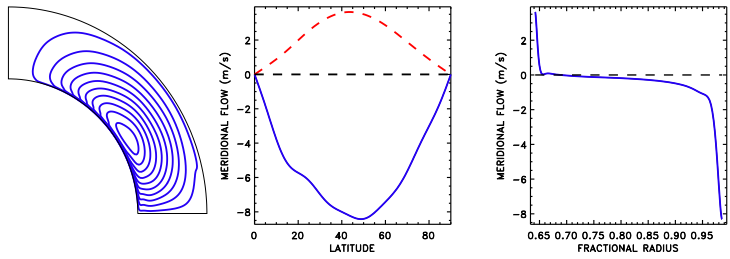}
\caption{Meridional flow of an M star with $M=0.6 M_\odot$ rotating with a period of 10 d. Left: isocontours of the stream function. Center: meridional flow speed at the top (blue solid line) and bottom (red dashed line). Positive values indicate flow towards the equator, negative towards the pole. Right: flow speed vs.~fractional stellar radius at mid-latitudes.
\label{m06_flow}
}
\end{center}
\end{figure}

The meridional flow is  driven by two effects, namely the gradient of the angular velocity parallel to the rotation axis, $\partial \Omega^2 / \partial z$, and the latitudinal gradient of the entropy, $\partial s / \partial \theta$.
Fig.~\ref{m06_flow} shows the flow pattern for the same case as Fig.~\ref{m06_dr}. The stream function $\psi$ is defined through
 $$
   \rho u_r = \frac{1}{r^2 \sin \theta}  \frac{\partial \psi}{\partial \theta}, \hspace{1cm}
   \rho u_\theta = \frac{1}{r \sin \theta} \frac{\partial \psi}{\partial r},
 $$
  where $\rho$ is the mass density and $u_r$ and $u_\theta$ the vertical and horizontal components of the meridional flow.
 Its isocontours are stream lines of the meridional flow. The spacing of the isolines, however, does not directly correspond to the flow speed, as is illustrated by the right panels in the figure. There are a thin layers of fast flow at both the top and bottom boundaries while the flow is much slower in the bulk of the convection zone. Note that the amplitude of the return flow at the bottom is about half that of the surface flow.

\section{Cycle times}
With the meridional flows computed above, we can now estimate the cycle times for a flux transport dynamo. We define the travel time,
$$
\tau = \frac{\pi  R_{\rm bot}}{2 u_m}
$$
where $R_{\rm bot}$ and $u_m$  radius and the average flow speed at the bottom of the convection zone. 
As one would have to integrate $1/u_\theta$ over latitude and that integral diverges when taken from zero to $\pi/2$, we define $u_m =0.4 \max(u_\theta)$ as the average flow speed. This choice reproduces the observed eleven year cycle time when applied to our model for the Sun and corresponds to integrating from $8^\circ$ to $82^\circ$. 

\begin{figure}
\begin{center}
\includegraphics[width=6.7cm]{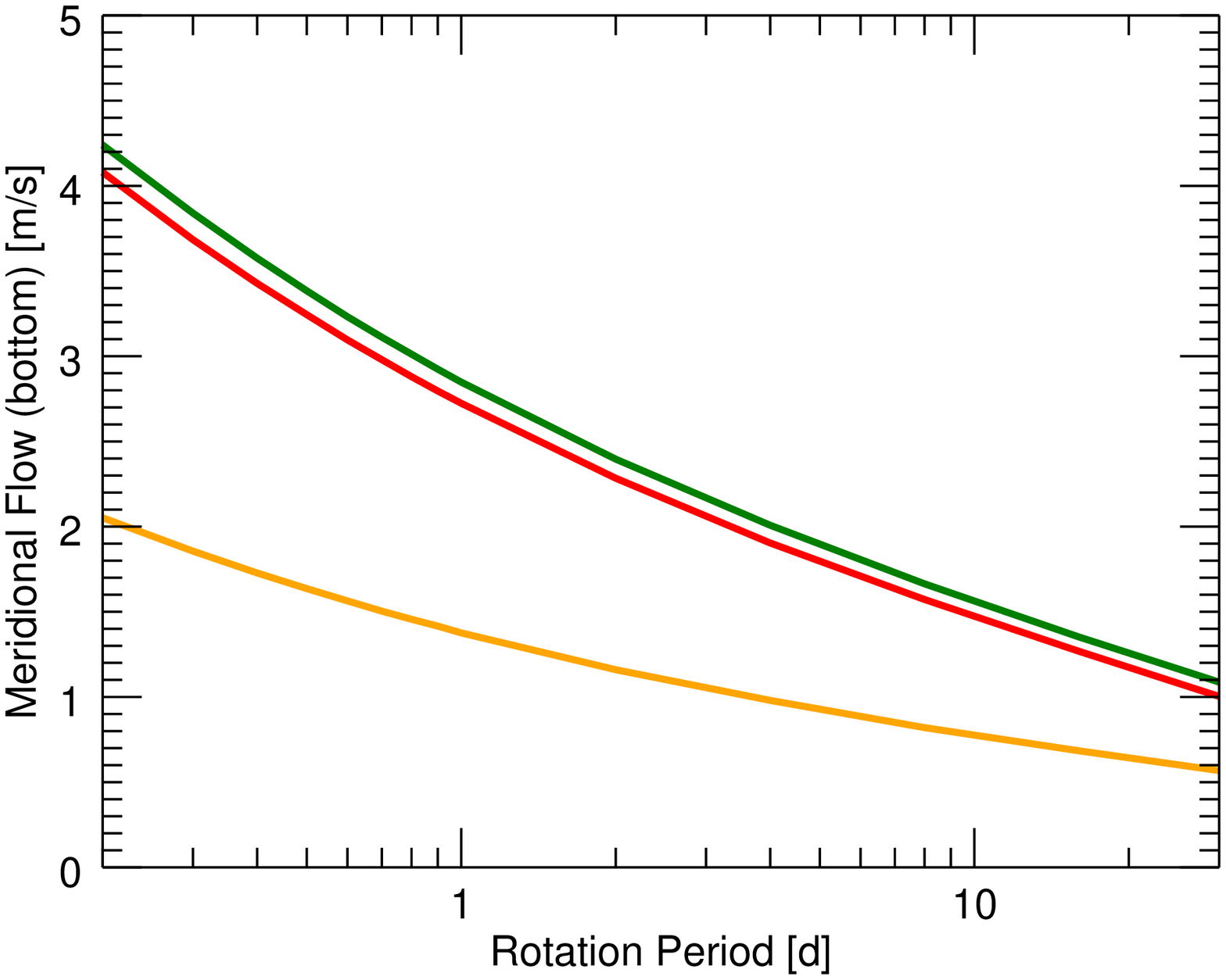}
\includegraphics[width=6.7cm]{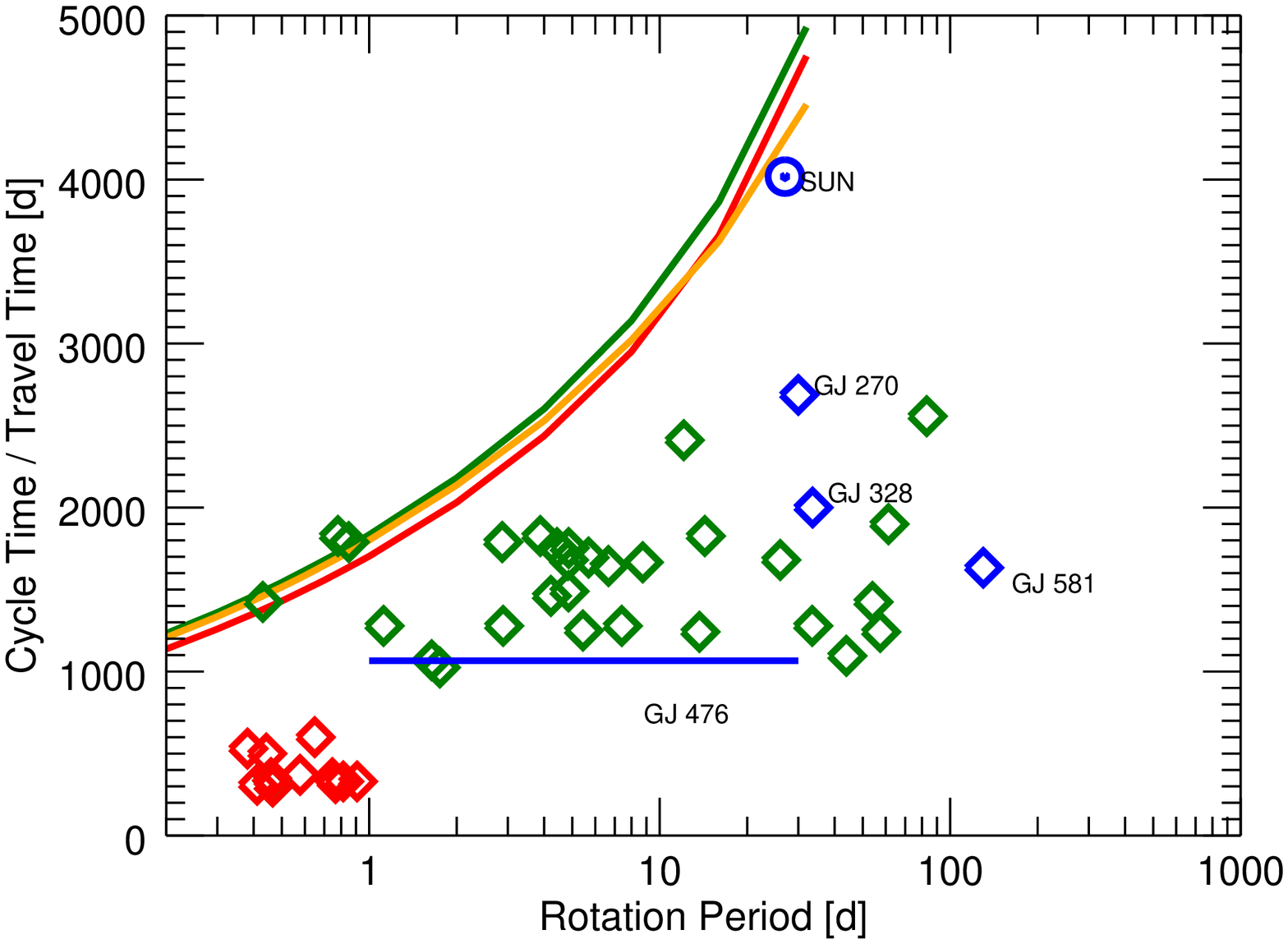}
\end{center}
\caption{Left: Meridional flow speed (left) and the resulting travel time at the bottom of the convection zone for stars of $M=0.66$~M$_{\odot}$ (green line), $M=0.60$~M$_{\odot}$ (red line),  and $M=0.40$~M$_{\odot}$ (yellow line). The right panel also shows the observed rotation periods and cycle times from Fig.~\ref{result} (diamods) and the Sun ($\odot$).}
\label{botflow}
\end{figure}

The left panel in Fig.~\ref{botflow} shows $u_m$  for all three models for a range of rotation periods from 0.2 to 30 days. For each model, the variation with the rotation period is rather moderate. The lines of 0.66 $M_\odot$ and 0.6 $M_\odot$ models are almost identical while the 0.4 $M_\odot$ for the model with 0.4 $M_\odot$ the flow amplitude is about half that of the models with larger masses. All cases show an increase of the flow speed with increasing rotation rate. 

The right panel of Fig.~\ref{botflow} shows the travel times computed with the flow speeds shown in the left panel and the observed cycle times from Fig.~\ref{result}. Despite the differences in flow speed, particularly between the $0.4 M_\odot$ case and the $0.6 M_\odot$ and $0.66 M_\odot$ models, the three curves are practically identical, i.e. the differences in flow speed and radius cancel. The sun is also very close to the three curves, despite being much more massive and luminous. 

A comparison between the observed cycle times and our model travel times shows that while there are three cases where both are the same, the vast majority of the stars in the sample shows cycle times that are substantially shorter than the travel times. The discrepancy is particularly large for the rapidly rotating stars observed by {\sc Kepler}, but also for some of the slow rotators. 

\section{Conclusions}
In the flux transport dynamo, the flow speed that sets the cycle time. As the gas in the solar convection zone is not a perfect conductor, the magnetic field must actually be expected to move a bit more slowly than the gas. That makes our estimated travel time a lower estimate for the cycle time, i.e.~all stars in Fig.~\ref{botflow} should lie on or above the green, yellow, and red lines. As almost all the stars in the sample lie below the lines, our findings do not support the flux transport dynamo as the mechanism behind the activity of these stars. 
%\clearpage
%
%
%\begin{figure}
%\begin{center}
%\includegraphics[width=8.0cm]{tflow.ps}
%\end{center}
%\caption{Travel times for stars of $M=0.66$~M$_{\odot}$ (green line), $M=0.60$~M$_{\odot}$ (red line),  and $M=0.40$~M$_{\odot}$ (yellow line) and observed cycle times. The Sun has been included for comparison.}
%\label{tflow}
%\end{figure}

\bibliography{iaus354}

\end{document}